# SmartEdge: Smart Healthcare End-to-End Integrated Edge and Cloud Computing System for Diabetes Prediction Enabled by Ensemble Machine Learning

Alain Hennebelle
*Independent Researcher in Information Technology Systems and Artificial Intelligence Applications*
Abu Dhabi, United Arab Emirates

Qifan Dieng
*Cloud Computing and Distributed Systems (CLOUDS) Lab, School of Computing and Information Systems*
University of Melbourne, Melbourne, Australia

Leila Ismail
[1]*Intelligent Distributed Computing and Systems (INDUCE) Lab, Department of Computer Science and Software Engineering, College of Information Technology*
[2]*Emirates Center for Mobility Research*
[3]*National Water and Energy Center*
UAE University, Al-Ain, Abu Dhabi, United Arab Emirates
Correspondence: leila@uaeu.ac.ae

Rajkumar Buyya
*Cloud Computing and Distributed Systems (CLOUDS) Lab, School of Computing and Information Systems*
University of Melbourne, Melbourne, Australia

*Abstract* -- The Internet of Things (IoT) revolutionizes smart city domains such as healthcare, transportation, industry, and education. The Internet of Medical Things (IoMT) is gaining prominence, particularly in smart hospitals and Remote Patient Monitoring (RPM). The vast volume of data generated by IoMT devices should be analyzed in real-time for health surveillance, prognosis, and prediction of diseases. Current approaches relying on Cloud computing to provide the necessary computing and storage capabilities do not scale for these latency-sensitive applications. Edge computing emerges as a solution by bringing cloud services closer to IoMT devices. This paper introduces SmartEdge, an AI-powered smart healthcare end-to-end integrated edge and cloud computing system for diabetes prediction. This work addresses latency concerns and demonstrates the efficacy of edge resources in healthcare applications within an end-to-end system. The system leverages various risk factors for diabetes prediction. We propose an Edge and Cloud-enabled framework to deploy the proposed diabetes prediction models on various configurations using edge nodes and main cloud servers. Performance metrics are evaluated using, latency, accuracy, and response time. By using ensemble machine learning voting algorithms we can improve the prediction accuracy by 5% versus a single model prediction.

*Keywords* — *Artificial Intelligence, Cloud Computing Diabetes, Diagnosis, Edge Computing, Ensemble Learning, Health care, eHealth, Internet of Things, Machine Learning, Prediction, Prognosis*

I. INTRODUCTION

The pervasive influence of Internet of Things (IoT) devices is significantly reshaping applications in different smart city domains such as healthcare, transportation, industry, and education [1]. Projections indicate that the IoT paradigm is poised to yield a staggering economic impact of $11 trillion annually, with an estimated deployment of one trillion IoT devices by 2025 [2]. These devices are anticipated to play a pivotal role in monitoring and managing diverse smart systems in real-time across different domains, including the realm of smart healthcare [3]. Within the context of smart hospitals, the deployment of Internet of Medical Things (IoMT) devices is becoming increasingly widespread. The IoMT devices often fail to satisfy the computation and communication requirements of healthcare applications due to low computing capabilities and limited battery life. While cloud-based approaches are designed to handle large volumes of data and process these data in complex compute-intensive applications, they result in high latency due to the long distance between the IoMT devices and the cloud. To remedy this latency issue, edge and fog computing are introduced bringing the cloud services closer to the IoMT devices and users. Edge computing uses nodes, gateway, and routers to deliver services that can be both energy efficient and low latency. There is a need to develop an AI-powered smart healthcare system that leverages the capabilities of integrated Edge-cloud computing. This work aims to address this void. We propose SmartEdge, an end-to-end automated system that supports low latency applications and provides high accuracy using ensemble learning diabetes prediction. Diabetes is one of the deadliest diseases in the world. It is then crucial to be able to detect it at an early stage [4]. SmartEdge provides a novel architecture for smart healthcare application enablement. It integrates the power of high accuracy ensemble-learning and low latency of Edge computing nodes, harnessing FogBus2 [5] as a backend framework. For diabetes prediction, SmartEdge provides ensemble learning with the most performing machine learning algorithms Random Forest, and the mostly used Logistic Regression, and Support Vector Machines[6]. Furthermore, SmartEdge emphasizes effective feature selection techniques, identifying and prioritizing key variables that play a pivotal role in diabetes prediction using real-life diabetes datasets. SmartEdge for diabetes prediction is designed to support a large number of connected patients, thanks to its powerful and flexible framework. Our main contributions are as follows:

- We proposed a system architecture that leverages ensemble prediction on the Edge.
- We developed SmartEdge, an automated end-to-end diabetes prediction system powered by ensemble

learning and the networking capacity of Edge-enabled infrastructure.
- We deployed SmartEdge using Fogbus2 in an IoT-edge-cloud integrated computing system.
- We evaluated SmartEdge performance using a variety of metrics in terms of system performance and model accuracy.
- We analyzed the ensemble learning approaches in SmartEdge using a set of different metrics: Accuracy F-measure, network latency, and response times. We used real-life patient data to determine the prevalence of type II diabetes.

The rest of this paper is organized as follows. We present the related work in Section 2, the background cloud/edge technologies description is provided in Section 3. The proposed architecture is described in Section 4, Section 5 details the design, and Section 6 the implementation of this architecture. In Section 7 we present the experimental setup and analyze the results of performance evaluation. Conclusions and future work are presented in Section 8.

## II. RELATED WORK

In the literature we can find a variety of approaches and innovations in the field of diabetes prediction using machine learning (ML), each contributing to some extent to understanding and presenting a methodology for tackling Type 2 Diabetes Mellitus (T2DM) through predictive analytics. While papers like [7] focus on specific populations, such as the Pima Indians, offering insights into the utility of ML in addressing diabetes within ethnically and geographically distinct groups,[8] work on a rural Chinese population underscores the importance of considering diverse demographic and geographic contexts in diabetes prediction, highlighting differences that may exist between rural and urban populations. The various studies utilize a range of ML models, including logistic regression, neural networks, decision trees, and ensemble methods, to predict the onset of Type 2 Diabetes. The work from [9] uses logistic regression as a baseline or comparison model, given its popularity for binary classification tasks such as predicting the presence or absence of diabetes. Logistic regression is favored for its interpretability and efficiency, making it a staple in medical research for assessing risk factors. Random forest is an ensemble method that combines multiple decision trees to improve prediction accuracy and robustness used in studies like [7] for classifying Pima Indian Diabetes Mellitus data. Decision trees are described as intuitive and easy to interpret, making them useful for understanding which features (e.g., glucose levels, BMI) are most predictive of diabetes. Work like [10] presents a way to use patient networks for disease prediction that leverages algorithms capable of handling relational data and graph-based representations, such as Graph Neural Networks (GNNs), to exploit the structure and relationships within patient data for improved prediction outcomes. Other common ML algorithms used for machine learning prediction include deep learning models, such as neural networks, that are employed in some of the papers to handle complex interactions between features and to model non-linear relationships. SVMs are used for example in [11] for diabetes prediction, the authors pinpoint their effectiveness in high-dimensional spaces and their ability to use different kernel functions to transform the feature space into one where a linear separator can divide the classes. Neural networks also are used, they are naturally powerful for high-dimensional data and can capture complex patterns that simpler models might miss, making them suitable for datasets with a rich set of patient features [12]. In work in [13], by leveraging a voting classification scheme, the authors underscore the value of ensemble methods in healthcare analytics, particularly in improving the accuracy and reliability of predictive models. Additionally, understanding which health-related features (e.g., BMI, blood pressure, glucose levels) significantly impact diabetes prediction is a common theme. [14] , for instance, investigate health-related features and their roles in enhancing prediction accuracy. [10] introduce an approach by using patient network-based models for disease prediction, suggesting the value of integrating diverse data sources and considering patient interactions or similarities. None of this work uses ensemble voting in an end-to-end prediction system implemented using the full scale of the edge/fog/cloud paradigm. This paper addresses the gap.

Table 1 summarizes the works on machine learning-based diabetes prediction.

## III. SMART EDGE HEALTHCARE SYSTEM ARCHITECTURE

The SmartEdge model uses an IoT Edge Cloud model to integrate software and hardware components to deliver accurate and fast diabetes prediction. SmartEdge is composed of different pieces of hardware equipment and software components. Fig.1 presents the architecture of SmartEdge. The different components and their interaction are explained next.

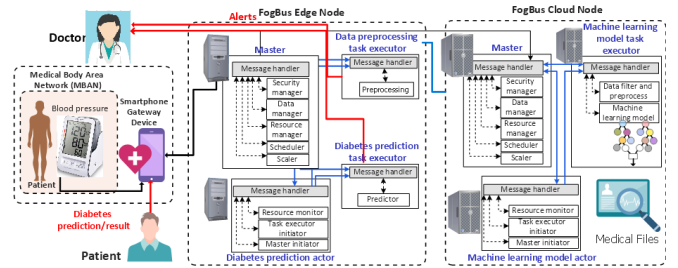

Fig. 1. : Smart Edge Healthcare System Architecture

### A. SmartEdge Hardware Components

1) Patient Monitoring devices: The patient can use several devices which aim to monitor diabetes risk factors [18]. [17] provides a survey on diabetes IoMT devices.

2) In SmartEdge this component constitutes a Blood Pressure (BP) monitor which can provide measures for blood pressure systolic/diastolic and heartbeat levels [17]. A smartphone using Bluetooth technology acts as the gateway device.

3) Gateway Device: IoMT devices monitoring patients send data into the system via gateway devices. This gateway is paired with its corresponding monitoring device for communication. Each patient owns a gateway device (e.g. smartphone, tablet, or laptop). The gateway device is the initiator of the processing.

4) Fogbus2 Framework Modules: there are 3 levels of modules in Fogbus2 that we use for SmartEdge.

TABLE I. RELATED WORK ON DIABETES PREDICTION

| Work | Dataset | Algorithms compared | Outperforming algorithm | IoT and Edge | Performance Analysis | Ensemble Voting | Distributed ML | AI Model Accuracy | End-to-end system |
|---|---|---|---|---|---|---|---|---|---|
| [14] | Private dataset from hospitals in Saudi Arabia | LR, SVM, DT, RF, and EMV | N/R | | √ | | | √ | |
| [15] | Private dataset from Slovenian primary healthcare institutions | LR, Glmnet, RF, XGBoost, and light GBM | LR | | √ | | | √ | |
| [16] | Private dataset from Hanaro Medical Foundation in Seoul, South Korea | LR, RF, SVM, XGBoost†, stacking†, soft voting†, and confusion matrix-based ensemble† | RF | | √ | | | √ | |
| [12] | Dataset 1: cross-sectional diabetes survey in Saudi Arabia; Dataset 2: national Health and Nutrition Examination Survey; Dataset 3: PIMA Indian | Bayes Point Machine (BPM), AP, DF, LD-SVM, DJ, boosted DT, and NN | DF | | √ | | | √ | |
| [9] | PIMA Indian | LR and DT | LR | | √ | | | √ | |
| [7] | PIMA Indian | NB, RF, and DT | RF | | √ | | | √ | |
| [8] | Henan rural cohort study | LR, CART, ANN, SVM, RF, and GBM | GBM | | √ | | | √ | |
| [10] | CBHS health funds company in Australia | LR, kNN, SVM, NM, DT, RF, XGBoost, and ANN | RF | | √ | | | √ | |
| [17] | Dataset 1: PIMA Indian; Dataset 2: Sylhet; Dataset 3: MIMIC III | LR, RF and SVM | RF | √ | √ | | | √ | √ |
| [11] | 1999–2020 NHANES database | CATBoost XGBoost RF, LR, SVM | CATBoost | | √ | | | √ | |
| This paper | PIMA Indian | CATBoost XGBoost RF, LR, SVM, Ensemble Voting | Ensemble Voting | √ | √ | √ | √ | √ | √ |

a) Master node: this component receives the requests from the gateway device and oversees assigning available resources in the Fogbus2 network to fulfill the request.
b) Worker Node: this component performs the tasks transmitted by the master node. A worker node receives data from the gateway and sends back the results. Such a node typically performs calculation tasks like data preprocessing and even more complex machine learning prediction.
c) Task Executor: this component carries out the functions assigned by the actor component. It includes a diabetes prediction task, which features a predictor module. This module utilizes the established prediction model to estimate diabetes occurrence in a user, drawing from data provided by a hypertension-monitoring IoT device.

5) Cloud servers: in the situation where edge resources are over-utilized, or some tasks receive data whose size exceeds their capacity, SmartEdge can leverage the Cloud servers. The tasks must be latency-tolerant. So, compute-intensive tasks like training machine learning models with updated datasets can be performed by HeathEdge, without compromising the stability of the system.

The system is designed for dynamic load balancing and efficient utilization of resources based on the current load conditions and the presence or absence of Cloud resources.

*B. Software Components*

The SmartEdge system is composed of the following software components. On the master node, the software component receives the job requests and/or input data from Gateway devices. The request input module receives job requests from Gateway devices just before transferring the data.

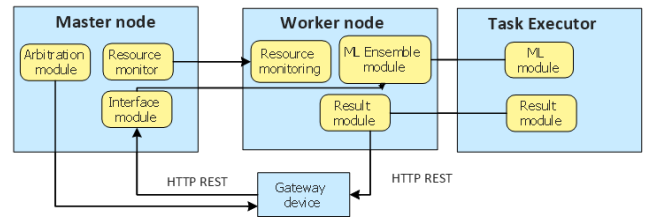

Fig. 2. SmartEdge Implementation

The Security Management module provides secure communication between different components and protects the collected data from unauthorized access or malicious tampering of data to improve system credibility and data integrity. The arbitration module (part of Resource Manager in the master node) takes as input the load statistics of all worker nodes and decides which node or subset of nodes to send jobs to in real-time. On the worker node: the software components perform the main tasks allocated by the Resource Manager of the master node. Worker nodes can comprise embedded devices or small computers. In SmartEdge, Worker nodes can contain sophisticated deep-learning models to process and analyze the input data and generate results. In addition, the Worker node can include other components for data processing, data filtering, and mining, Big Data analytics, and storage. Task Executor node is similar to the Worker node, in the specific implementation for SmartEdge, the Task Executor performs individual machine learning prediction.

Gateway Device: The Gateway device initiates the job request process by sending job requests and input data to the Broker node.

### C. Processing

For data processing in the non-cloud scenario, the Gateway device sends the job to the Worker/Broker node. The Broker node, based on the workers' loads, cpu, and memory usage decides which node is more suitable. The decision-making process may involve considerations of compromised workers and security measures. The efficiency of load checks and decision-making processes is discussed concerning the increasing number of Workers.

### D. System workflow

System workflow is described in Figure 3. The process begins with Gateway devices initiating job requests and data transfer to the Broker node. The Broker node receives job requests and input data from Gateway devices. Security Management ensures secure communication and protects data integrity. The Arbitration module in the Resource Manager dynamically allocates tasks to Worker nodes based on load statistics. Worker nodes perform assigned tasks, which can involve sophisticated machine-learning models. Worker nodes accommodate diverse functions, including data processing, filtering, mining, analytics, and storage.

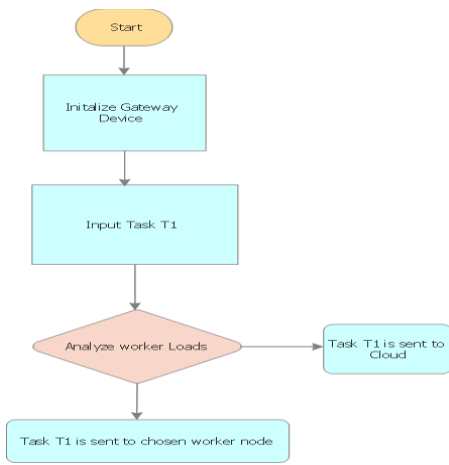

Fig. 3. Health Edge Scheduling

## IV. SMARTEDGE DESIGN

### A. Data Preprocessing

We use a Pima Indians Diabetes Database dataset with relevant features for diabetes prediction.

Data preprocessing involves addressing missing values, eliminating outliers, scaling data, and selecting relevant features. Missing values can be managed by either deleting the corresponding records or inserting synthetic values. These synthetic values can be created using either statistical methods (such as mean, mode, or median) or machine learning techniques (such as SMOTE [19]). Data scaling can be performed through normalization and/or standardization. It is essential to normalize numerical features with different ranges to prevent model bias towards features with larger ranges [91]. Recursive Feature Elimination [20], which identifies the most relevant features for predicting diabetes.

### B. Machine Learning

The system architecture of SmartEdge described in the previous section uses a diabetes dataset that includes risk factors and diabetes data for patients. The dataset is then used to develop a prediction model. For a new user, the system takes diabetes risk factors data, measured using sensors and medical tests, as input and uses the prediction model to determine whether the user will have diabetes or not, and sends back the prediction results. This is implemented using the preprocessing module, machine learning module, and gateway interface as described in this section.

Diabetes dataset processing involves data splitting: we divide data into training, validation, and testing sets in the ratio of 70:10:20. The model training utilizes the training set for training the model, the validation set for tuning the model, and the test set for evaluating model performance on new data.

### C. Communication

The process begins with the Gateway sending a Job request to the Broker node in every scenario. Broker Node Response (Worker Node or Cloud): Based on the scenario, the Broker node responds to the Gateway with either: Worker IP address (of the same LAN) or Master IP address (with/without cloud forwarding). In the Broker-only case, the Broker node may or may not check the loads of workers. If all workers have heavy loads or are compromised, and the Cloud is disabled, the Broker sends its IP without cloud forwarding to the Gateway. If there are workers not heavily loaded, the Broker sends the IP address of the least loaded Worker node to the Gateway. The decision-making process involves load checks on Worker nodes. Increasing the number of Workers would increase the arbitration time as more load checks need to be performed. In the non-cloud scenario, the Gateway device directly sends the job (input data for analysis) to the Worker/Broker node. The Worker/Broker node then runs pre-processing on the received input data for analysis.

The SmartEdge communication protocols ensure efficient utilization of resources by dynamically allocating tasks based on the scenario, load conditions, and user-defined configurations. The system is designed to adapt to varying conditions, promoting effective load balancing and secure data processing in different deployment scenarios.

### D. Algorithms

Voting algorithms: When using the ensemble method for machine learning, we use a voting algorithm. There are two main ways to perform the vote: hard voting and soft voting. The hard voting method is straightforward: each member of the classifier group casts a vote for a prediction, and the collective decision is based on which option receives the most votes. For instance, if two classifiers determine a diabetic diagnosis, then the diabetic diagnosis is selected. Soft voting considers the prediction confidence levels of each classifier. Every classifier provides a probability estimate for each category, and the ensemble's decision reflects the category with the greatest cumulative probability. For instance, if a classifier assigns a 0.9 probability to positive diabetics and another gives a 0.2 probability of not being, the collective prediction will favor the diabetic diagnosis.

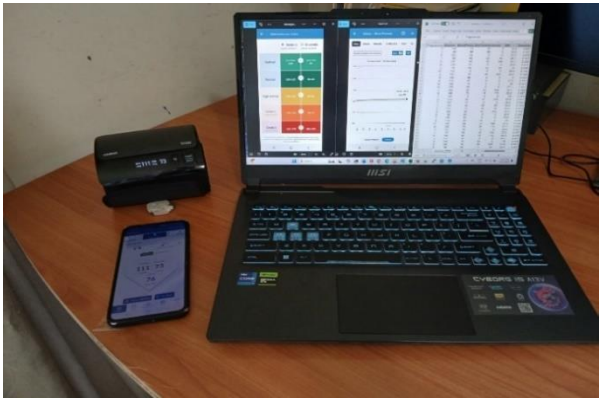

Fig. 4. SmartEdge Experimental Setup

## V. SYSTEM IMPLEMENTATION

The elements outlined in Section 5 are implemented using various programming languages. Python is utilized for the pre-processing and ensemble machine-learning components. The preprocessing module removes observations with missing risk factor values based on their distribution. For the ensemble machine learning application, the SciKit Learn Library [21] was employed. Our voting scheme utilized, RF, Cat Boost, and LR. During training, the model randomly distributed data among classifiers. During diagnosis, it considered all predicted classes and outputted the majority prediction. Worker selection followed the default Fogbus2 policy, considering the worker with the minimum CPU load. The chosen worker received the CSV file for analysis. The Execution Interface Module in each worker instantiated the Ensemble Learning code for data analysis. The result was sent back to the Worker node, which, in turn, forwarded the ensembled result to the gateway device. The system can sustain continuous monitoring thanks to its robust framework infrastructure.

## VI. PERFORMANCE EVALUATION

We implement SmartEdge for ensemble learning diabetes prediction using Fogbus2 framework [5] on the backend. We evaluate the performance of the proposed system in terms of prediction accuracy, communication and computation times, and network overhead. We also assess the performance improvement when using voting ensemble methods for diabetes prediction versus single model prediction.

### A. Experimental Environment

We conduct the experiments on an edge-cloud setup consisting of 3 cloud servers and 3 edge nodes. Each cloud server node is a virtual machine with 2 CPU AMD EPYC 7763 64-Core Processor, 8GB RAM. This CPU features data L1 cache of 32 KiB per core, 8-way set associative, L1 instruction cache of 32 KiB per core, 8-way set associative, L2 cache of 512 KiB per core, 8-way set associative and L3 cache of 256 MiB. Each Edge node is a Raspberry PI. All the nodes are connected through a WIFI network. The connection between the edge and the cloud nodes is a 1Gbits network.

### B. Dataset

We employ the Pima Indian Diabetes dataset [22] from the National Institute of Diabetes and Digestive and Kidney Diseases. This dataset serves to diagnose whether a patient is diabetic or non-diabetic based on different risk factors. The dataset encompasses nine features: (1) pregnancies (number of times pregnant), (2) glucose (plasma glucose concentration at 2 hours in an oral Glucose Tolerance Test), (3) diastolic blood pressure (mm Hg), (4) triceps skin fold thickness (mm), (5) 2-hour serum insulin (µU/ml), (6) Body Mass Index (BMI) calculated as weight in kg divided by the square of height in meters, (7) diabetes pedigree function (indicating the expected genetic influence of diabetic and non-diabetic relatives), (8) age (in years), and (9) outcome (diagnosis of diabetes with a value of 1 for diabetic and 0 for non-diabetic).

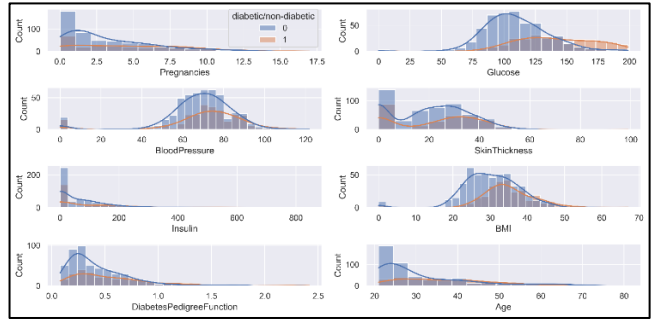

Fig. 5. Distribution of PIMA Indian diabetes dataset features for diabetic and non-diabetic classes.

The dataset comprises medical records for 768 patients (i.e., observations), with 268 (34.9%) identified as diabetic and 500 (65.1%) as non-diabetic. Figure 5 presents the distribution of numerical features for diabetic and non-diabetic classes. To preprocess the data, we eliminate observations with missing values. As depicted in Figure 1, skin thickness, blood pressure, and BMI features have '0' values, indicating missing data. Consequently, the preprocessed dataset after removing observations with missing values consists of 537 records, including 179 (33.3%) diabetics and 358 (66.7%) non-diabetics. The preprocessed information is stored in a Comma Separated Value (.csv) file.

### C. Experiments

We use the PIMA Indian dataset to measure the performance of our ensemble learning model in predicting the prevalence of diabetes. The dataset is partitioned into 70%, 10%, and 20% for training, validation, and testing of the model respectively. The preprocessed dataset consists of 537 observations. The dataset is divided equally among the different worker nodes to train prediction models. The system is evaluated in terms of accuracy, recall, precision, F-measure, ROC, AUC, arbitration time, latency (communication time), execution time, network response time, and latency. We compare the performance of our ensemble model with individual learning algorithms used in our ensemble approach. The list of experiments is as follows: Perform training of the model on the chosen dataset. Perform the prediction using the trained models and collect performance measurement using the following metrics: prediction accuracy, precision, recall, F-measure, ROC, and AUC, for both settings: single-model prediction, and ensemble model prediction. We perform learning using the following models SVM, RF, Logical Regression, and Decision Tree. We then perform the prediction and record the values of the voting result, and Accuracy, ROC, F-measure, and AUC performance metrics, according to the number of edge nodes. We also evaluate the confidence in the model. The latency and response time are also measured for the 5 cases. In addition, we also record network usage execution time in different cases. The time measurement is done using gettimeofday function on Linux system. We first make a run with all voting entities on the same machine to analyze the performance of voting. Then make different experiments varying the number of edges nodes and roles assignment. The experimental scenarios are:

1. 4 nodes: A+BCD, ABCD are all R Pi model 5, 8 cpu cores, 4 gb ram, arm,,connected with WiFi.
- Node A: FogBus2 User
- Node B: FogBus2 Master + Actor0

- Node C: FogBus2 Actor1
- Node D: FogBus2 Actor2

2. 2 nodes: A+B: A and B are R Pi model 5, 8 cpu cores, 4 gb ram, arm, WiFi
- Node A: FogBus2 User
- Node B: FogBus2 Master + Actor0 + Actor1+ Actor2

3. 4 nodes with cloud A+BCD: A is a RP i model 5, 8 cpu cores, 4 gb ram, arm, WiFi, and BCD are AWS EC2 m4.4xlarge, 16 cpu cores, 64 gb ram.
- Node A: FogBus2 User
- Node B: F ogBus2 Master + Actor0
- Node C: FogBus2 Actor1
- Node D: FogBus2 Actor2

### D. Performance Results Analysis

We test our system as described in the experiments and perform the analysis. The analysis is done using prediction accuracy, precision, recall, F-measure, ROC, and AUC and it is prediction-based. We evaluate the performance of the ensemble prediction versus the centralized one, in terms of accuracy, and latency. The numerical parameters for this study are the response time, latency, and execution time.

When studying the accuracy performance in Table III and Fig 7, we can see that the chosen prediction models perform similarly on the selected dataset. We notice that using ensemble voting model improves consistently accuracy by 5%.

### E. Response times study

Figure 10, 11, and 12 illustrate the response time at the Broker node for various scenarios, including: (1) A+BCD, (2) A+B, (3) A+BCD Cloud. Our results show that the response time is minimal (approximately 55 ms) when tasks are sent directly to the Broker/Master or Cloud. However, as the number of Edge nodes increases (scenario A+BCD compared to A+B), the Broker needs to perform additional load checking at each Worker node to identify the node with the minimum load, resulting in a corresponding increase in response time.

### F. Latency study

Figures 8,9,10 illustrate the variation in latency, which is the sum of communication time and queuing delay. (We observe that the latency is similar when tasks are sent to the Broker or any of the Edge nodes, as all communication occurs through single-hop data transfers. In contrast, the latency is significantly higher in the Cloud for the user to master communication as there are multiple hops to reach the cloud. But for internal communication in the cloud, latency is lower because it uses a fast-wired network, and the edge nodes (in A+BCD) scenario use a WIFI network. When studying the accuracy performance in Table III and Fig 7, we can see that the chosen prediction models perform similarly on the selected dataset. We notice that using ensemble voting model improves consistently accuracy by 5%. Figure 13 illustrates the variation in execution time. As expected, the execution time is very low in the Cloud setup due to the higher resource availability. The Broker's execution time is shorter than that of the Worker nodes, which are Raspberry Pis with lower clock frequency processors.

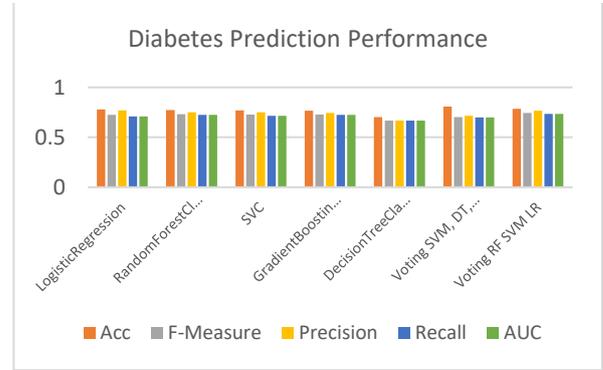

Fig. 6. Diabetes Prediction Performance

Notably, our results also demonstrate the benefits of ensemble voting. Regardless of the number of Edge nodes, the ensemble voting approach consistently outperforms the non-ensemble method (best or average), achieving higher accuracy.

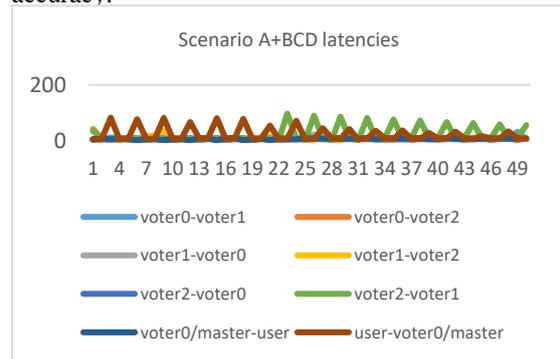

Fig. 7. Scenario A+BCD Latencies

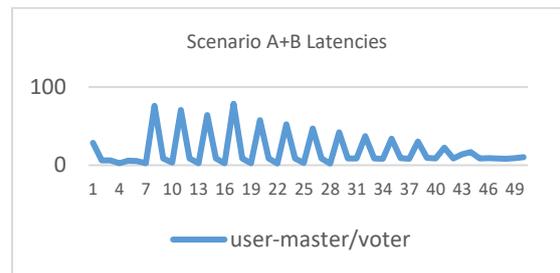

Fig. 8. Scenario A+B Latencies

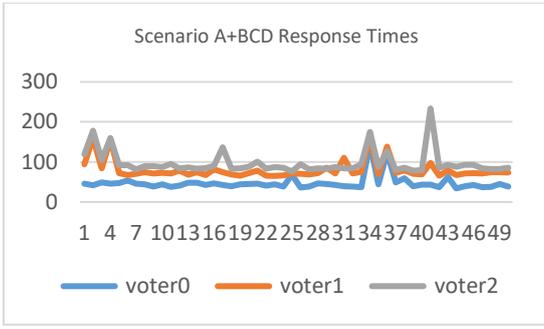

Fig. 9. Scenario A+BCD Response times

## G. Discussion

The application deployment system offers flexible configurations that cater to different user requirements, providing a trade-off between accuracy and latency. Based on our experimental results, we recommend the following deployment strategies for SmartEdge:

- For heavy and latency-tolerant tasks, the cloud configuration is recommended. This setup is necessary to ensure successful task completion, as resource-constrained Edge worker nodes may struggle to handle such tasks otherwise.
- For latency-critical tasks, worker nodes are the ideal choice. This setup ensures rapid result delivery due to the proximity of worker nodes. If network bandwidth constraints exist, ensemble machine learning predictions should be disabled to conserve resources. However, if resources permit, enabling ensemble voting can improve accuracy.

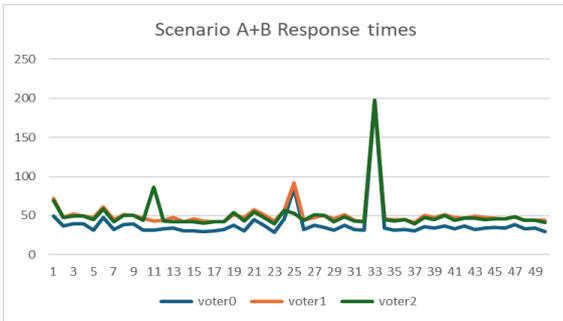

Fig. 10. Scenario A+B Response times

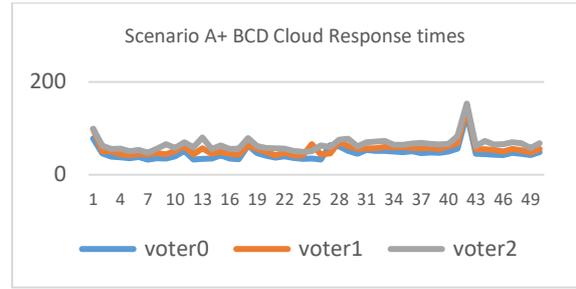

Fig. 11. Scenario A+ Cloud BCD Response times

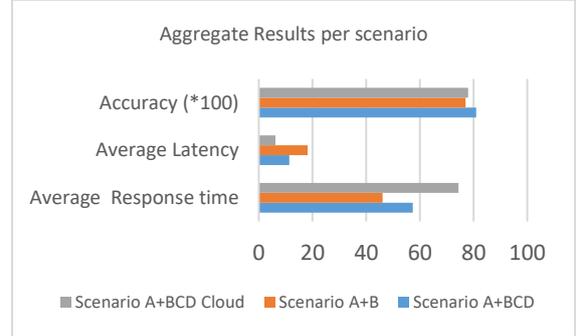

Fig. 12. Aggregate results according to scenario

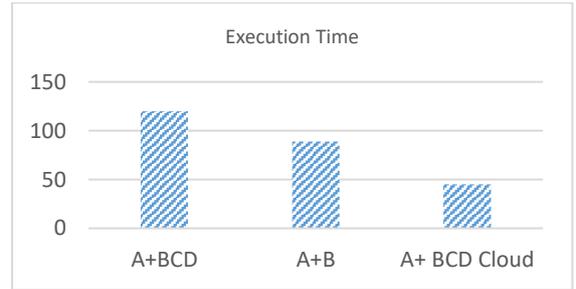

Fig. 13. Execution time according to scenario.

TABLE II. DIABETES PREDICTION RESULTS

| Algorithm | Acc | F-Measure | Precision | Recall | AUC |
|---|---|---|---|---|---|
| LogisticRegression | 0.7784 | 0.7243 | 0.7685 | 0.7102 | 0.7102 |
| RandomForestClassifier | 0.7722 | 0.7325 | 0.7507 | 0.725 | 0.725 |
| GradientBoostingClassifier | 0.7667 | 0.7293 | 0.7436 | 0.7241 | 0.7241 |
| DecisionTreeClassifier | 0.7037 | 0.6661 | 0.6679 | 0.6681 | 0.6681 |
| Voting SVM, DT, LR | 0.8086 | 0.7021 | 0.7154 | 0.6981 | 0.6981 |
| Voting RF SVM LR | 0.784 | 0.7435 | 0.7664 | 0.7333 | 0.7333 |

## VII. Conclusions and Future Work

In this research, our focus was specifically on improving healthcare services for diabetes prediction through the development of an Edge-based Smart Healthcare System, named SmartEdge, that utilizes deep learning and IoT for the automated prediction of diabetes. SmartEdge improves healthcare delivery by offering edge services that adeptly handle data from diverse IoT devices, enhancing patient data management. By incorporating deep learning within edge computing devices SmartEdge has been applied successfully to real-world heart disease analysis. Unlike previous efforts that lacked the use of deep learning and thus suffered from inadequate prediction accuracy, our approach enables the embedding of complex machine learning networks within edge computing frameworks. We achieved this through innovative communication and model distribution strategies, such as ensembling, ensuring high precision with minimal latency. The efficacy of SmartEdge was demonstrated through validation against real patient data, employing the FogBus framework within an edge computing environment, and evaluating the system's performance across several metrics including power consumption, network response time, latency, and accuracy.

Looking ahead, we aim to evolve SmartEdge by integrating cost-effective execution strategies that consider various Quality of Service (QoS) attributes and edge-cloud cost dynamics. Although SmartEdge currently operates on file-based input data, future iterations will seek direct sensor data integration to enhance user interaction. Additionally, we plan to refine the model training approach, which presently involves independent training at each worker node followed by an ensemble of models through bagging. By exploring more sophisticated ensemble techniques, we anticipate further enhancements in prediction accuracy. Moreover, we intend to expand the robustness and versatility of the proposed architecture to support a wider range of edge computing applications beyond healthcare, including agriculture, weather forecasting, traffic management, and smart city initiatives.

## Acknowledgment

This research was funded by the National Water and Energy Center of the United Arab Emirates University (Grant 31R215).